\documentclass[conference,a4paper]{IEEEtran}
 \usepackage[utf8]{inputenc}
\usepackage{bbm,dsfont,mathrsfs,fixmath,amsthm}
\usepackage{amsmath,amssymb,graphicx} 
\usepackage{amssymb}
\usepackage{graphicx}
\usepackage{cite}
\usepackage{comment}
\usepackage{epsfig}
\usepackage{color}
\usepackage{graphicx}
\usepackage[nolist,printonlyused]{acronym}
\usepackage{ctable}
\newcommand{\gTX}{g_{\mathrm{tx}}}
\newcommand{\gRX}{g_{\mathrm{rx}}}
\newcommand{\Xnm}{X\left[n,m\right]}
\newcommand{\XnmPrime}{X\left[n',m'\right]}
\newcommand{\HnmPrime}{H_{n,m}\left[n',m'\right]}
\newcommand{\nPrimeSum}{\sum_{n'=0}^{N-1}}
\newcommand{\mPrimeSum}{\sum_{m'=0}^{M-1}}
\newcommand{\kPrimeSum}{\sum_{k'=0}^{N-1}}
\newcommand{\lPrimeSum}{\sum_{l'=0}^{M-1}}
\newcommand{\nSum}{\sum_{n'=0}^{N-1}}
\newcommand{\mSum}{\sum_{m'=0}^{M-1}}
\newcommand{\pSum}{\sum_{p=0}^{P-1}}
\newcommand{\PsiMat}{\boldsymbol{\Psi}^p_{k,k'}\left[l,l'\right]}
\newcommand{\Pavg}{P_{\mathrm{avg}}}
\newcommand{\rect}{{\rm rect}}

\newtheorem{theorem}{Theorem}
\newtheorem{lemma}[theorem]{Lemma}%

\begin{document}

\begin{acronym}
	\acro{AWGN}{additive white Gaussian noise}
	\acro{OTFS}{orthogonal time frequency space}
	\acro{ISFFT}{inverse symplectic finite Fourier transform}
	\acro{SFFT}{symplectic finite Fourier transform}
	\acro{CRLB}{Cram\'er-Rao lower bound}
	\acro{ISI}{inter-symbol interference}
	\acro{ICI}{inter-carrier interference}
	\acro{SNR}{signal-to-noise ratio}
	\acro{MAE}{mean absolute error}
	\acro{OFDM}{orthogonal frequency division multiplexing}
	\acro{CP}{cyclic prefix}
	\acro{DFT}{discrete Fourier transform}
	\acro{IDFT}{inverse discrete Fourier transform}
	\acro{ML}{maximum likelihood}
	\acro{MMSE}{minimum mean square error}
	\acro{FMCW}{frequency modulated continuous wave}
	\acro{MSE}{mean square error}
\end{acronym}

\title{Performance Analysis of Joint Radar and Communication using OFDM and OTFS}

\author{
\IEEEauthorblockN{Lorenzo Gaudio$^{1}$, Mari Kobayashi$^{2}$, Bj\"orn Bissinger$^{2}$, Giuseppe Caire$^{3}$ }
\IEEEauthorblockA{$^{1}$University of Parma, Italy \\
$^{2}$Technical University of Munich, Munich, Germany \\
$^{3}$Technical University of Berlin, Germany\\ 
Emails: lorenzo.gaudio@studenti.unipr.it, \{mari.kobayashi, bjoern.bissinger\}@tum.de,~caire@tu-berlin.de }}

\maketitle

\begin{abstract}
We consider a joint radar estimation and communication system using orthogonal frequency division multiplexing (OFDM) and orthogonal time frequency space (OTFS) modulations. 
The scenario is motivated by vehicular applications where a vehicle equipped with a mono-static radar wishes to communicate data to its target receiver, while estimating parameters of interest related to this receiver. 
By focusing on the case of a single target, we derive the maximum likelihood (ML) estimator and the Cram\'er-Rao lower bound on joint velocity and range estimation. Numerical examples demonstrate that both digital modulation formats can achieve as accurate range/velocity estimation as state-of-the-art radar waveforms such as frequency modulated continuous wave (FMCW) while sending digital information at their full achievable rate. We conclude that it is possible to obtain significant data transmission rate without compromising the radar estimation capabilities of the system.

\end{abstract}

\section{Introduction}
The key-enabler of high-mobility networks is the ability to continuously track the dynamically changing environment (state) and react accordingly by exchanging information with each other. The high cost of spectrum and hardware will inevitably encourage that both state estimation and communication shall be operated by sharing the same frequency bands. Towards emerging applications such as vehicular to everything (V2X), we consider a joint radar and communication system where a radar equipped transmitter (vehicle) wishes to estimate the parameters of a target receiver and simultaneously send data to this receiver, as already investigated in the literature (see \cite{sturm2011waveform,kumari2018ieee, nguyen2017delay} and references therein). Although most of existing works build on a resource-sharing approach such that time or frequency resources are split into either radar estimation or data communication \cite{kumari2018ieee, nguyen2017delay}, a synergetic design can potentially yield a significant performance gain, as demonstrated in an information theoretic framework \cite{kobayashi2018joint}. Motivated by this result, we study the performance of a joint radar and communication system using two digital modulation formats, namely, the well-known \ac{OFDM} and the recently proposed \ac{OTFS} (see \cite{raviteja2018interference} and references therein).   

By focusing on a single-target case, we characterize the joint radar and communication performance in terms of \ac{MSE} and achievable rate. More precisely,  the \ac{ML} estimator and the \ac{CRLB} on velocity and range estimation are derived. Our numerical examples, inspired by the parameters of IEEE 802.11p, demonstrate that both waveforms provide similar radar performance while \ac{OTFS} yields better multiplexing gain than \ac{OFDM}. Surprisingly, the two waveforms perform similarly as \ac{FMCW}, one of the typical automotive radar waveforms \cite{patole2017automotive}, while achieving a significant communication rate for free. 

It is worth noticing that \ac{OFDM} has been extensively studied for radar estimation (see \cite{sturm2011waveform,braun2014ofdm,nguyen2017delay} and references therein). However, none of these works has addressed explicitly the joint performance in terms of MSE and achievable communication rate. 
A comparison of radar estimation between \ac{OTFS} and \ac{OFDM} has been independently developed in a very recent work \cite{viterboOTFSradar}. By neglecting part of the fractional Doppler and delay shifts and further focusing on a low-complexity matched filter approach for \ac{OTFS}, this work concluded that \ac{OTFS} has some advantages over \ac{OFDM} since it can handle larger Doppler shifts and longer range. In contrast, the current work considers \ac{ML} estimator without neglecting the fractional part of Doppler and delay shifts, while restricting to Doppler shifts of practical relevance for automotive applications, such that the \ac{ICI} incurred by \ac{OFDM} is definitely negligible. Under these conditions, we obtain different conclusions as \cite{viterboOTFSradar}. 

The paper is organized as follows. In section \ref{sec:phy-model} we present the physical model. In section \ref{sec:OFDM} and \ref{sec:OTFS}, we derive the \ac{ML} estimator and the 
\ac{CRLB} for \ac{OFDM} and \ac{OTFS}, respectively. Section \ref{sec:sim-results} provides numerical results and Section \ref{sec:conclusions} concludes the paper.

\section{Physical model}\label{sec:phy-model}
We consider a joint radar and communication system over a total bandwidth of $B$ [Hz] operating at the carrier frequency $f_c$ [Hz]. We assume that a transmit vehicle, equipped with a mono-static full-duplex radar, wishes to convey a message to its target receiver while estimating parameters related to the same receiver. 
For simplicity, we focus on a point target model, so that the link between a transmitter and a receiver can be represented by as a single line-of-sight channel \cite{li2007mimo,kumari2018ieee}. In a multi-carrier system, the total bandwidth is divided into $M$ subcarriers, i.e., $B=M \Delta f$, where $\Delta f$ [Hz] denotes the subcarrier bandwidth. 
For a given maximum delay $\tau_{\max}$ and a given maximum Doppler shift $\nu_{\max}$, the symbol duration $T$ and the subcarrier spacing $\Delta f$ are chosen to satisfy
\begin{equation}\label{eq:max-delay-Doppler}
	\nu_{\max}  < \Delta f\,,
	\quad
	\tau_{\max} < T\,,
\end{equation}
with $T=1/\Delta f$. 
Suppose there are $P$ targets and the $p$-th target vehicle is at the relative range of $r_p$ [m] and is moving at the relative velocity of $v_p$ [m/sec] with respect to the transmitter. We model the radar channel as a $P$-tap time frequency selective channel given by 
\begin{equation}\label{eq:radarchannel-tau-nu}
	h(t, \tau) = \sum_{p=0}^{P-1} h_p \delta(\tau-\tau_p) e^{j2\pi \nu_p t}\,,
\end{equation}
where $h_p$ is the complex channel gain, $\nu_p = \frac{2v_p f_c}{c}$ and $\tau_p=\frac{2 r_p}{c}$ denotes a round-trip Doppler shift and delay, respectively. 
By taking into account the one-way Doppler shift and delay, the forward communication channel is given by 
\begin{align} \label{eq:h_com}
 h_{\rm com} (t, \tau) = {g_0} e^{j \pi \nu_0 t} \delta\left(\tau- \frac{\tau_0}{2}\right)\,,
\end{align}
where $g_0$ denotes the complex channel gain. In both \ac{OFDM} and \ac{OTFS}, data symbols $x_{n,m}$, for $n=0,\dots,N-1$ and $m=0,\dots,M-1$, are arranged in an $N\times M$ grid $\Lambda$, and satisfy 
the average power constraint, i.e.
\begin{equation}\label{eq:avg-power}
	\frac{1}{NM}\sum_{n=0}^{N-1}\sum_{m=0}^{M-1}\mathbb{E}\left[\left|x_{n,m}\right|^2\right]\leq\Pavg\,.
\end{equation}

\section{OFDM}\label{sec:OFDM}
\subsection{Input Output Relation}
Consider a standard \ac{OFDM} modulation with \ac{CP} in order to avoid the \ac{ISI}. The resulting \ac{OFDM} symbol duration is $T_o= T_{\rm cp} +T$, where $T_{\rm cp}$ and $T$ denote the \ac{CP} and data symbols duration, respectively. 
Provided the maximum delay $\tau_{\max}$ we typically choose $T_{\rm cp}=C\frac{T}{M}$, with $C=\lceil \frac{\tau_{\max}}{T/M}\rceil$, where $\lceil\cdot\rceil$ is the rounding-up operation. The \ac{OFDM} frame duration is thus $T_f^{\rm ofdm}=NT_o$. The continuous-time \ac{OFDM} transmitted signal with \ac{CP} is given by 
\begin{equation}
	s(t) =  \sum_{n=0}^{N-1}\sum_{m=0}^{M-1}  x_{n, m}\rect(t - n T_o)e^{j 2\pi m \Delta f (t-T_{\rm cp}-nT_o)}\,,
\end{equation}
where $\rect(t)$ is one for $t\in [0, T_o]$ and zero otherwise. 
Ignoring the noise, the signal received after the time-frequency selective channel \eqref{eq:radarchannel-tau-nu} is
\begin{equation}\label{eq:OFDM-RxSignal}
	y'(t) = \int h(t,  \tau) s(t-\tau) d\tau = \sum_{p=0}^{P-1} h_ps(t-\tau_p)e^{j2\pi \nu_p t}\,.
\end{equation}
By sampling every $\frac{T}{M}$ and removing the CP in each OFDM symbol, we obtain
\begin{align}
	y'_{n, m} &=  y'(t) |_{t=n T_o + T_{\rm cp}+ m{T}/{M}}=\sum_{p=0}^{P-1}h_p e^{j2\pi n T_o \nu_p }\nonumber\\
	&\sum_{m'=0}^{M-1}  x_{n, m'}  e^{j 2\pi \frac{m}{M} \left( \frac{\nu_p}{\Delta f} + m'  \right)}e^{-j 2\pi {m'} \Delta f \tau_p}\,.
\end{align}
Applying the \ac{DFT} and using the orthogonal property, the output is given by 
\begin{align}\label{eq:in-out-ofdm}
	y_{n, m} &= \frac{1}{M}\sum_{i=0}^{M-1} y'_{n, i} e^{-j2\pi \frac{mi}{M}}\nonumber\\
	& \approx\sum_{p=0}^{P-1}h_p e^{j2\pi n T_o \nu_p}  e^{-j 2\pi {m} \Delta f \tau_p} x_{n, m}\,,
\end{align}
where the approximation follows by letting $\nu_{\max}\ll\Delta f$\footnote{Note that this approximation can be justified in a number of 
scenarios. For example, consider a scenario inspired by IEEE 802.11p with $f_c=5.89$ GHz and the subcarrier spacing $\Delta f=156.25$ KHz. 
This yields $v_{\max} \ll 14325$ [km/h], which is reasonable even for a relative speed of 400 [km/h]. The same holds for IEEE 802.11ad with $f_c$=60 GHz and $\Delta f $= 5.15625 MHz \cite{cordeiro2010ieee}.}. 
Under the approximated channel input-output relation \eqref{eq:in-out-ofdm}, it readily follows that the Doppler shift and the delay are decoupled, which makes joint range and velocity estimation simple
 (see e.g. \cite{braun2014ofdm,sturm2011waveform}). 
By focusing for simplicity on a single-target case ($P=1$), we neglect the $p$-path subscript from now on. Since data symbols are known by the radar receiver (the transmitter itself), and the noise is i.i.d. Gaussian circularly symmetric, the radar receiver can undo the data symbol phase without changing the noise statistics. Therefore, the radar observation after this symbol-by-symbol phase rotations can be written as
\begin{equation}\label{eq:z-ofdm}
	z_{n,m}=A_{n,m}h e^{j2\pi nT_o\nu}e^{-j2\pi m\Delta f\tau}+w_{n,m}\,,
\end{equation}
where $A_{n,m}=\left|x_{n,m}\right|$ denotes the amplitude of the transmitted symbol and $w_{n, m}$ is \ac{AWGN} with zero mean and unit variance. 

\subsection{Maximum Likelihood Estimator} 
We derive the \ac{ML} estimator of channel gain/range/velocity for the observation model in \eqref{eq:z-ofdm} by generalizing \cite[Chapter 3.3.3]{braun2014ofdm} to the case of arbitrarily amplitude $\boldsymbol{A}$, with $\boldsymbol{A} =\{A_{n, m}\}$. For the set of parameters $\boldsymbol{\theta}=(h, \nu, \tau)$, 
we wish to find the estimator minimizing the log-likelihood function
\begin{equation}\label{eq:ll}
	l(\boldsymbol{z}|\boldsymbol{\theta}, \boldsymbol{A} ) = \sum_n\sum_m \left|z_{n, m}\hspace{-0.07cm}-\hspace{-0.03cm}h A_{n, m}e^{j2\pi\left(\nu nT_o-m\Delta f\tau\right)}\right|^2\hspace{-0.15cm}.\hspace{-0.06cm}
\end{equation}
Assuming $(\nu, \tau)$ known, by setting the derivative of $l(\boldsymbol{z}|\boldsymbol{\theta}, \boldsymbol{A} )$ with respect to $h$ equal to zero, we obtain the estimator $\hat{h}$ of the complex channel gain $h$, which is
\begin{equation}\label{eq:hest}
	\hat{h} =\frac{Z(\nu, \tau)}{\sum_{n, m} A^2_{n, m}}\,,
\end{equation}
where we defined a \ac{DFT}/\ac{IDFT} operation as
\begin{align}
	Z(\nu, \tau) \triangleq \sum_{m=0}^{M-1}\sum_{n=0}^{N-1} z_{n,m}A_{n, m}e^{-j2\pi \nu n T_o }  e^{j 2\pi {m} \Delta f \tau},
\end{align}
which is a two-dimensional periodogram. By plugging \eqref{eq:hest} into \eqref{eq:ll} and following similar steps as \cite[Chapter 7.2.2]{richards2014fundamentals}), we obtain the estimator
\begin{equation}\label{eq:241}
	(\hat{\nu}, \hat{\tau})=\arg\max_{(\nu, \tau)\in \Gamma}  \left|Z(\nu, \tau)\right|^2\,,
\end{equation}
where we considered a discretized set $\Gamma$ of delay and Doppler frequency axes with step sizes ${1}/\left({M'\Delta f}\right)$ and ${1}/\left({N'T_{\rm o}}\right)$, respectively, with $N'\geq N$ and $M'\geq M$.

In summary, to compute the \ac{ML} estimator of $(h, \tau, \nu)$ the following steps are done:
\begin{enumerate}
	\item Compute the \ac{DFT}/\ac{IDFT} output $Z(\nu, \tau)$.
	\item Choose $(\hat{\nu}, \hat{\tau})$ maximizing $\left|Z(\nu, \tau)\right|^2$  over $\Gamma$. 
	\item Let the channel gain be $\hat{h}= {Z(\hat{\nu}, \hat{\tau})}/\left(\sum_{n, m} A_{n, m}^2\right) $.
\end{enumerate}
The resulting velocity and range radar estimations are given by 
$\hat{v} = \frac{\hat{\nu}c}{2f_c}$ and $\hat{r} = \frac{\hat{\tau}c}{2}$.

\subsection{CRLB}
\newcommand{\Id}{{\bf I}}
\newcommand{\st}{\mathsf{t}}
\newcommand{\thetav}{\hbox{\boldmath$\theta$}}

Consider the vector of unknown $\thetav=(\alpha, \varphi, f, \st)$, where $\alpha = |h|$, $\varphi = \angle(h)$, $f=T_{o} \nu$, and $\st= \Delta f \tau$, from \eqref{eq:z-ofdm} we obtain
\begin{align}\label{eq:OFDMSingleTarget}
	z_{n, m}= A_{n, m} \alpha e^{j \varphi} e^{j2\pi  n f }  e^{-j 2\pi {m} \st} +  w_{n,m}\,.
\end{align}
By letting $s_{n, m}=A_{n, m} \alpha e^{j \varphi} e^{j2\pi  n f }  e^{-j 2\pi {m} \st}$, we derive the $4\times 4$ Fisher information matrix defined as 
\begin{align}\label{eq:fisher-matrix}
	[\Id(\thetav,\boldsymbol{A})]_{i, j} 
	&= 2 \Pavg  \Re\left\{\sum_{n,m} \left[\frac{\partial s_{n,m}}{\partial \theta_i}\right]^*\left[\frac{\partial s_{n,m}}{\partial \theta_j}\right]\right\}.
\end{align}
After straightforward algebra, we are able to prove the following result.
\begin{lemma}
In the regime of large $M$ and $N$, the CRLB of $f$ and $\st$ are given by  
\begin{subequations}
\begin{align}
\label{eq:Varf2}
\sigma_{\rm \hat{f}}^2 
	&\geq 
	 \frac{ 6  }{ |h|^2 \Pavg(2\pi)^2 M N ( {N}^2 - 1 ) }\,,
\\
\label{eq:Vart2}
\sigma_{\rm \hat{\st}}^2 
	&\geq  
	  \frac{ 6  }{|h|^2 \Pavg(2\pi)^2 M N ( {M}^2 - 1 ) }\,.
\end{align}
\end{subequations} 
\end{lemma}
For a special case of constant envelope ($A_{n, m} = \sqrt{\Pavg}$ for all $n, m$), the above expressions coincide with those in \cite[Section 3.3]{braun2014ofdm}.  

\section{OTFS}\label{sec:OTFS}
\subsection{Input Output Relation}
The  transmitter first applies the \ac{ISFFT} to represent data symbols $\{x_{k,l}\}$ in the time-frequency domain, i.e., $\Xnm$, then generates the continuous time signal
\begin{equation}
	s\left(t\right)=\hspace{-0.1cm}\sum_{n=0}^{N-1}\sum_{m=0}^{M-1}\Xnm\gTX\left(t-nT\right)e^{j2\pi m\Delta f\left(t-nT\right)},\,\hspace{-0.18cm}
\end{equation}
where $\gTX$ denotes a transmit pulse. 
The \ac{OTFS} frame duration is thus $T_f^{\mathrm{otfs}}=NT$. The noiseless received signal $r\left(t\right)$, after the channel in \eqref{eq:radarchannel-tau-nu}, is given in  \eqref{eq:OFDM-RxSignal}.
 Defining the cross ambiguity function between two pulses as in \cite{matz2013time}
\begin{equation}
	C_{u,v}\left(\tau,\nu\right)\triangleq\int_{-\infty}^{\infty}u\left(t\right)v^*\left(t-\tau\right)e^{-j2\pi\nu t}dt\,,
\end{equation}
the output of the matched filter is given by 
\begin{equation}
	Y\left(t,f\right)=C_{r,\gRX}\left(t,f\right)=\int r\left(t'\right)\gRX^*\left(t'-t\right)e^{-j2\pi ft'}dt'\,.
\end{equation}
By sampling at $t=nT$ and $f=m\Delta f$, the received samples in the time-frequency domain are given by 
\begin{align}\label{eq:Y-time-frequency}
	Y\left[n,m\right]&=Y\left(t,f\right)|_{t=nT,f=m\Delta f}\nonumber\\
	&=\nPrimeSum\mPrimeSum\XnmPrime\HnmPrime\,,
\end{align}
where, by letting $h_p'\triangleq h_pe^{j2\pi\nu_p\tau_p}$, we have
\begin{align}\label{eq:363}
	H_{n,m}&\left[n',m'\right]\triangleq\pSum h_p'e^{j2\pi n'T\nu_p}e^{-j2\pi m\Delta f\tau_p}\nonumber\\
	&C_{\gTX,\gRX}\left(\left(n-n'\right)T-\tau_p,\left(m-m'\right)\Delta f-\nu_p\right).
\end{align}
Finally, we obtain the received samples in the Doppler-delay domain applying the \ac{SFFT} to \eqref{eq:Y-time-frequency}, i.e.
\begin{align}\label{eq:y-sampled}
	y\left[k,l\right]&=\frac{1}{NM}\nSum\mSum Y\left[n,m\right]e^{-j2\pi\left(\frac{nk}{N}-\frac{ml}{M}\right)}\nonumber\\
	&=\kPrimeSum\lPrimeSum x_{k',l'} g_{k,k'}\left[l,l'\right]\,,
\end{align}
where the cross-talk channel of the Doppler-delay couple $\left[k',l'\right]$ seen by $\left[k,l\right]$ is given by
\begin{equation}
	g_{k,k'}\left[l,l'\right]=\pSum h_p'\PsiMat\,,
\end{equation}
with the channel matrix $\PsiMat$ defined in \eqref{eq:Psi}.
\begin{figure*}[t]
\begin{equation}\label{eq:Psi}
	\PsiMat=\frac{1}{NM}\sum_{n,n',m,m'}e^{j2\pi n'T\nu_p}e^{-j2\pi m\Delta\tau_p}e^{-j2\pi\left(\frac{nk}{N}-\frac{ml}{M}\right)}C_{\gRX,\gTX}\left(\left(n-n'\right)T-\tau_p,\left(m-m'\right)\Delta f+\nu_p\right)\,.
\end{equation}
\vspace{-0.7cm}
\end{figure*}
By stacking the $N\times M$ matrices of transmitted symbols and received samples to column vectors of length $NM$, we obtain the vector input-output relation as
\begin{equation}\label{eq:y}
	\boldsymbol{y}=\pSum h_p'\boldsymbol{\Psi}^p\boldsymbol{x}+\boldsymbol{w}\,,
\end{equation}
where $\boldsymbol{\Psi}^p$ is the $NM\times NM$ matrix obtained from \eqref{eq:Psi}, $\boldsymbol{w}$ denotes the AWGN with 
zero mean and identity covariance. Notice that our input-output relation in \eqref{eq:y} is exact and holds for any pair of transmit/receive pulses. 

Letting $\gTX\left(t\right)$ and $\gRX\left(t\right)$ be rectangular pulses of length $T$, it readily follows that the cross-ambiguity function is non-zero only for $n'=n$ and for $n'=n-1$ since 
the maximum channel delay $\tau_{\mathrm{max}}<T$.  
For further derivation, we consider also the approximated cross-ambiguity function given by 
\begin{align}\label{eq:cross-ambiguity}
	C_{\gRX,\gTX}\left(\tau,\nu\right)&=\int_0^T\gTX\left(t\right)\gRX^*\left(t-\tau\right)e^{-j2\pi\nu t}dt\nonumber\\
	&\approx \frac{1}{M}  \sum_{i=0}^{M-1-l_{\tau}} \exp\left(j 2\pi \nu \frac{Ti}{M}\right)\,,
\end{align}
where $l_{\tau}=\lceil \frac{\tau}{T/M}\rceil$ is an integer in $[0,M-1]$. The cross-talk matrix $\PsiMat$ using the approximated cross ambiguity function and rectangular pulses is given in \eqref{eq:Psi-final}.
\begin{figure*}[t]
	\begin{equation}\label{eq:Psi-final}\
	\begin{aligned}
	\PsiMat
	&\approx \frac{1}{NM}\frac{1-e^{j2\pi\left(k'-k+\nu_pNT\right)}}{1-e^{j2\pi\frac{\left(k'-k+\nu_pNT\right)}{N}}}\frac{1-e^{j2\pi\left(l'-l+\tau_pM\Delta f\right)}}{1-e^{j2\pi\frac{\left(l'-l+\tau_pM\Delta f\right)}{M}}}e^{j2\pi\nu_p\frac{l'}{M\Delta f}}
	\begin{cases}
	\begin{array}{ll}
	1 & l'\in\left[0,M-1-l_{\tau_p}\right] \\
	e^{-j2\pi\left(\frac{k'}{N}+\nu_pT\right)} & l'\in\left[M-l_{\tau_p},M-1\right] \\
	\end{array}\hspace{-0.13cm}.
	\end{cases}\hspace{-1cm}
	\end{aligned}
	\end{equation}
	\noindent\rule{\textwidth}{0.4pt}
\end{figure*}


\subsection{Maximum Likelihood Estimator}
By focusing on the single-target case $\left(P=1\right)$ and neglecting the $p$-path subscript, we wish to find the ML estimator for the set of unknown parameters $\boldsymbol{\theta}=\left(h',\tau,\nu\right)$. The log-likelihood function to be minimized is given by  
\begin{equation}
	l\left(\boldsymbol{y}|\boldsymbol{\theta},\boldsymbol{x}\right)=\left|\boldsymbol{y}-h'\boldsymbol{\Psi}\left(\tau,\nu\right)\right|\,,
\end{equation}
where symbols in $\boldsymbol{x}$ are known at the radar receiver. We now follow the same steps as for \ac{OFDM}. Assuming $\left(\tau,\nu\right)$ known, the estimator $\hat{h}'$ of the channel gain $h'$ is given by 
\begin{equation}\label{eq:h-estimator}
	\hat{h}'=\frac{\boldsymbol{x}^H\boldsymbol{\Psi}^H\boldsymbol{y}}{\boldsymbol{x}^H\boldsymbol{\Psi}^H\boldsymbol{\Psi}\boldsymbol{x}}\,,
\end{equation}
where $H$ indicates the transpose complex conjugate. We readily obtain the estimate $\hat{\tau},\hat{\nu}$ of $\tau,\nu$ using  \eqref{eq:h-estimator} as
\begin{align}
	\left(\hat{\tau},\hat{\nu}\right)
	&=\arg\max_{\left(\tau,\nu\right)\in\Gamma}\frac{\left|\boldsymbol{x}^H\boldsymbol{\Psi}\left(\tau,\nu\right)^H\boldsymbol{y}\right|^2}{\boldsymbol{x}^H\boldsymbol{\Psi}\left(\tau,\nu\right)^H\boldsymbol{\Psi}\left(\tau,\nu\right)\boldsymbol{x}}\,,
\end{align}
where $\Gamma$ is specified in \eqref{eq:241}. 

\subsection{CRLB}
In order to derive the \ac{CRLB}, we use the approximated channel matrix given in \eqref{eq:Psi-final}. Referring to the Fisher information matrix in  \eqref{eq:fisher-matrix}, we let
\begin{equation}
	s[k, l]=\sum_{k'=0}^{N-1}\sum_{l'=0}^{M-1}h'\boldsymbol{\Psi}_{k,k'}\left[l,l'\right]x\left[k',l'\right]\,.
\end{equation}
Note that the channel matrix contains the two unknown parameter $\tau$ and $\nu$.
In order to express the derivative, let us introduce the following indices
\begin{equation}
\begin{cases}
	l'_{\mathrm{ICI}}\triangleq l'\in\left[0,M-1-l_{\tau_p}\right]\\
	l'_{\mathrm{ISI}}\triangleq l'\in\left[M-l_{\tau_p},M-1\right]
\end{cases}.
\end{equation} 
The derivative w.r.t. $\tau$ is given by
\begin{align}
	&\frac{\partial\boldsymbol{\Psi}_{k,k'}\left[l,l'\right]}{\partial\tau}=\sum_n e^{j2\pi\left(\nu NT-k+k'\right)\frac{n}{N}}\sum_m e^{j2\pi\left(l-l'-\tau M\Delta f\right)}\nonumber\\
	&\left(-j2\pi m\Delta f\right)\frac{e^{j2\pi\nu\left(\frac{l'}{M\Delta f}\right)}}{NM}
\begin{cases}
\begin{array}{ll}
1 & l'_{\mathrm{ICI}}\\
e^{-j2\pi\left(\frac{k'}{N}+\nu T\right)} & l'_{\mathrm{ISI}}
\end{array}.
\end{cases}\hspace{-0.2cm}
\end{align}
The derivative w.r.t. $\nu$ is given by
\begin{align}
&\frac{\partial\boldsymbol{\Psi}_{k,k'}\left[l,l'\right]}{\partial\nu}=\frac{j2\pi}{NM}\sum_m e^{j2\pi\left(l-l'-\tau M\Delta f\right)}e^{j2\pi\nu\left(\frac{l'}{M\Delta f}\right)}\nonumber\\
&\Bigg[\sum_n e^{j2\pi\left(\nu NT-k+k'\right)\frac{n}{N}}
	\begin{cases}
	\begin{array}{ll}
		\frac{l'}{M\Delta f} & \hspace{-0.2cm} l'_{\mathrm{ICI}}\\
		e^{-j2\pi\left(\frac{k'}{N}+\nu T\right)}\left(\frac{l'}{M\Delta f}-T\right) & \hspace{-0.2cm} l'_{\mathrm{ISI}} \\
	\end{array}
	\end{cases}\nonumber\\
&+nT\sum_n e^{j2\pi\left(\nu NT-k+k'\right)\frac{n}{N}}
	\begin{cases}
	\begin{array}{ll}
		1 & l'_{\mathrm{ICI}}\\
		e^{-j2\pi\left(\frac{k'}{N}+\nu T\right)} & l'_{\mathrm{ISI}}\\
	\end{array}\Bigg].
	\end{cases}\hspace{-1cm}
\end{align}
The \ac{CRLB} expressions follow by applying the Fisher information matrix in \eqref{eq:fisher-matrix}.

\section{Simulation Results}\label{sec:sim-results}

\begin{table}
	\centering
	\renewcommand*{\arraystretch}{1.5}
	\caption{Simulation parameters}
	\begin{tabular}{|c|c|}
		\hline
		\multicolumn{2}{|c|}{IEEE 802.11p \cite{nguyen2017delay}}                                                                                                                                                       \\ \hline\hline
		$f_c=5.89$ GHz & $M=64$ \\ \hline
		$B=10$ MHz & $N=50$ \\ \hline\hline
		$\Delta f=B/M=156.25$ kHz & $T_{\mathrm{cp}}=\frac{1}{4}T=1.6\,\mu$s \\ \hline
		$T=1/\Delta f=6.4\,\mu s$ & $T_{\mathrm{o}}=T_{\mathrm{cp}}+T=8\,\mu$s \\ \hline
		$r^{\rm otfs}_{\mathrm{max}}<Tc/2\simeq960$ m & $r^{\rm ofdm}_{\mathrm{max}}<T_{\rm cp}c/2\simeq240$ m \\ \hline\hline
		$\sigma_{\mathrm{rcs}}=1$ m\textsuperscript{2} & $G=100$ \\ \hline
		$r=20$ m & $v=80$ km/h \\ \hline
	\end{tabular}
	\label{table:parameters}
\end{table}

\begin{figure}
	\centering
	\includegraphics[scale=0.555]{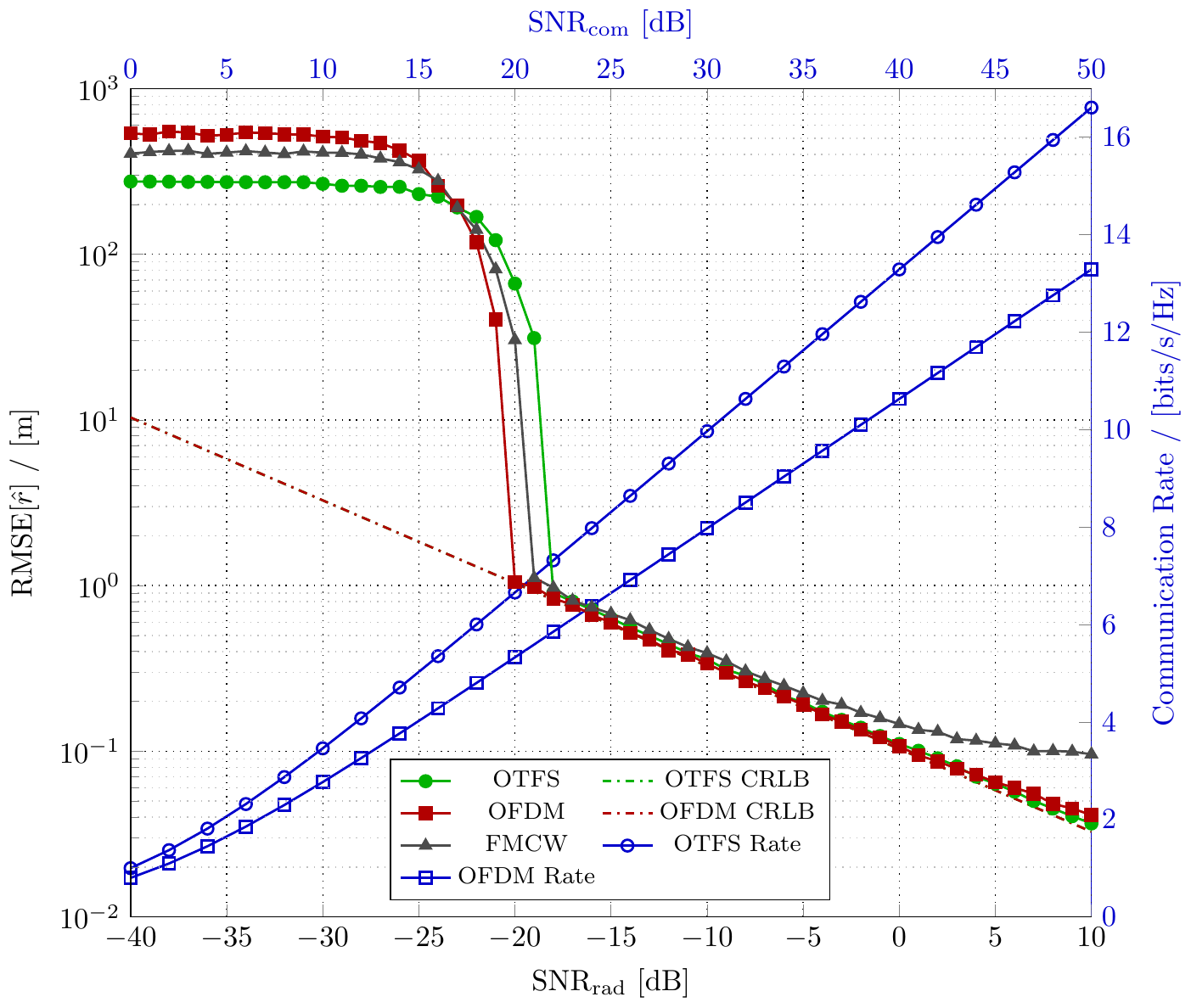}
	\caption{Root MSE (left y-axis) of the target range estimation $\hat{r}$ vs $\mathrm{SNR_{rad}}$ (bottom x-axis). The figure also shows the communication rate (right y-axis) vs $\mathrm{SNR_{com}}$ (top x-axis).
	}
	\label{fig:MAE-range}
\end{figure}
\begin{figure}
	\centering
	\includegraphics[scale=0.555]{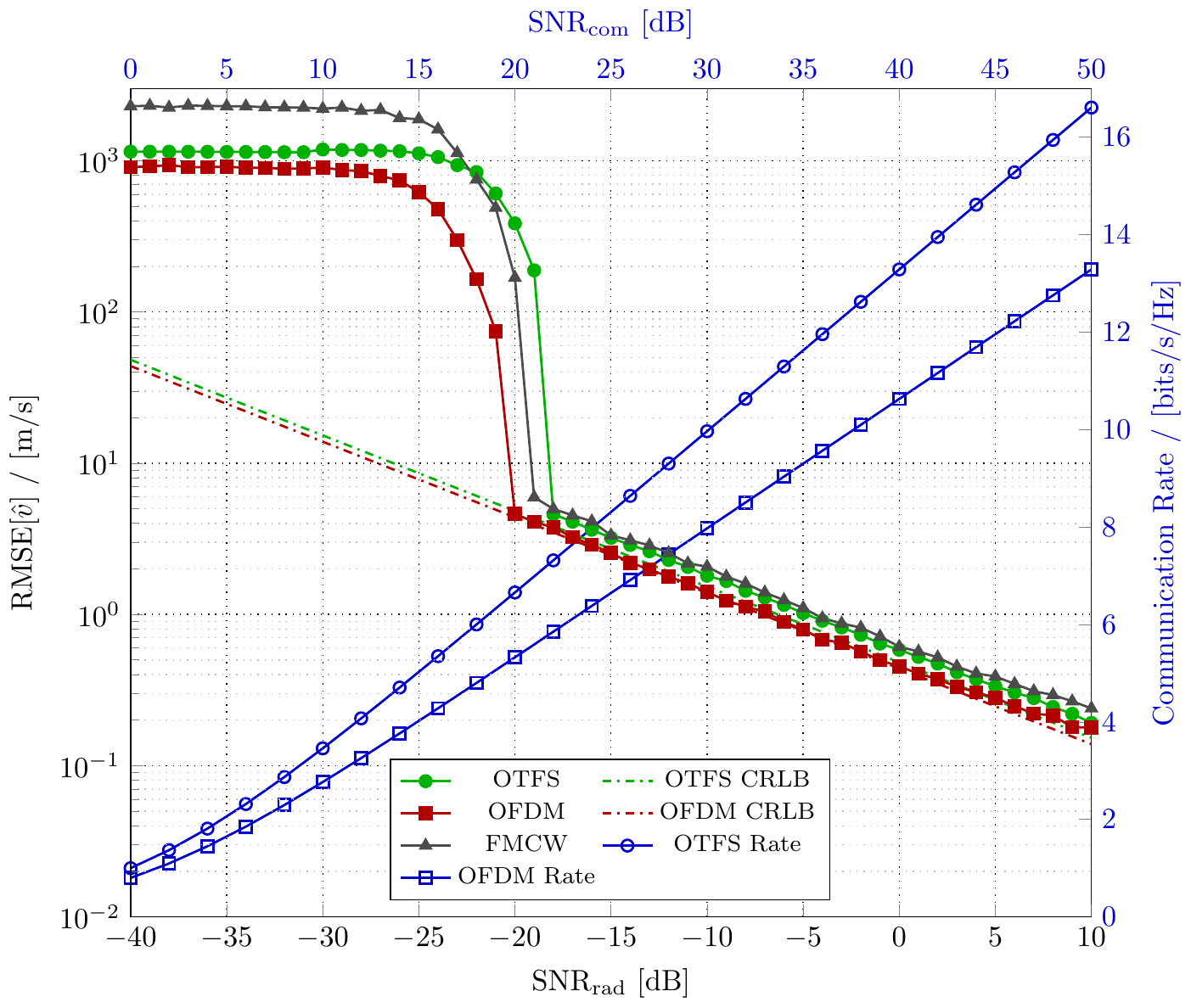}
	\caption{Root MSE (left y-axis) of the target velocity estimation $\hat{v}$ vs $\mathrm{SNR_{rad}}$ (bottom x-axis) and the communication rate as a function of $\mathrm{SNR_{com}}$ (top x-axis).
	}
	\label{fig:MAE-velocity}
\end{figure}

Simulation has been performed using the system model specified in Section \ref{sec:phy-model}.
Let the received radar and communication \ac{SNR} be
\begin{equation}
	\mathrm{SNR_{rad}}=\left|h_0\right|^2\Pavg\,,\quad\mathrm{SNR_{com}}=\left|g_0\right|^2\Pavg\,,
\end{equation}
with
\begin{equation}
	\left|h_0\right|=\sqrt{\frac{\lambda^2\sigma_{\mathrm{rcs}}G^2}{\left(4\pi\right)^3r^4}}\,
	,
	\quad\left|g_0\right|=\sqrt{\frac{\lambda^2G^2}{\left(4\pi\right)^2r^2}}\,,
\end{equation}
where $\lambda=c/f_c$ is the wavelength, $\sigma_{\mathrm{rcs}}$ is the radar cross section in $\mathrm{m}^2$, $G$ is the antenna gain, and $r$ is the distance between transmitter and receiver. In order to characterize the joint radar and communication performance, we provide the communication rate at the communication receiver for both \ac{OFDM} and \ac{OTFS}
\begin{subequations}\label{eq:57}
\begin{align}
	C_{\mathrm{OFDM}}&=\left(\frac{T}{T+T_{\mathrm{cp}}}\right)\log_2\left(1+\mathrm{SNR_{com}}\right)\,, \\
	C_{\mathrm{OTFS}}&=\frac{1}{NM}\log_2\det\left(\boldsymbol{I}+\mathrm{SNR_{com}}\boldsymbol{\Psi}\boldsymbol{\Psi}^H\right)\,.
\end{align}
\end{subequations}

Using the parameters listed in Table \ref{table:parameters}, we show the range estimation in terms of root \ac{MSE} (RMSE) and the communication rate \eqref{eq:57} for \ac{OFDM} and \ac{OTFS} in Figure \ref{fig:MAE-range}. 
Similarly, Figure \ref{fig:MAE-velocity} provides the velocity estimation and the communication rate. As a reference, we show the radar performance of \ac{FMCW}, as one of the popular automotive waveforms \cite{patole2017automotive}, using the same bandwidth and time resource. In both figures, we observe that the \ac{CRLB} of the three waveforms is almost identical.

First we observe that the radar performance is similar for three waveforms. It is remarkable that OFDM and OTFS, simultaneously sending data symbols, are able to achieve as accurate performance as \ac{FMCW}.
Second, we remark that \ac{OTFS} performs better than \ac{OFDM} in terms of communication rate by achieving a higher multiplexing gain. This is because \ac{OFDM} incurs an overhead due to \ac{CP}.  
It is worth noticing that \ac{OFDM} has additional constraints in terms of maximum range and velocity. Namely, the maximum delay is limited by the \ac{CP} duration, yielding  the maximum range $r_{\max}<cT_{\mathrm{cp}}/2$. Moreover, in order to ignore the \ac{ICI} as in \eqref{eq:z-ofdm}, the maximum Doppler shift must be significantly smaller than the subcarrier spacing $\Delta f$, yielding the maximum velocity $v_{\max} \ll \frac{c \Delta f}{2 f_c}$. However, the advantages of \ac{OTFS} in terms of estimation range limitations and achievable rate come at a considerable cost in complexity of the receiver, which implies a block-wise optimal decoder operating jointly on the whole block of symbols of size $MN$.

\section{Conclusions}\label{sec:conclusions}

In this paper, we analyzed the performance of a joint radar estimation and communication system based on OFDM and OTFS over the time frequency selective channel. Namely, we derived the ML estimator and the CRLB for both waveforms which enable us to compare them in terms of radar estimation MSE and communication rate. 
Although restricted to a simplified scenario with a single target, our numerical examples demonstrated that two waveforms provide as accurate radar estimation as \ac{FMCW} while providing a non-negligible communication rate for free. Our future works include the comparison with other radar waveforms, the extension to a multi-target case, and the performance analysis of \ac{OTFS} under more practical receivers.

\bibliography{IEEEabrv,tradeoff}

\end{document}